\documentclass[journal]{IEEEtran}

\usepackage{amsmath,amsfonts}
\usepackage{algorithmic}
\usepackage{algorithm}
\usepackage{array}
\usepackage[caption=false,font=normalsize,labelfont=sf,textfont=sf]{subfig}
\usepackage{textcomp}
\usepackage{stfloats}
\usepackage{url}
\usepackage{verbatim}
\usepackage{graphicx}
\usepackage{cite}
\usepackage{xcolor}
\usepackage{balance}
\usepackage{lastpage}
\usepackage{enumitem}
\usepackage{subfig}
\usepackage{bm} % para negrito em 
\usepackage{soul}   % para tachado
\usepackage{multirow}

\usepackage{tikz}
\definecolor{gold}{rgb}{0.85,.66,0}
\definecolor{viol}{rgb}{.68, .75, 1}

\begin{document}

\title{A Light Fixture Color Temperature and \\ Color Rendering Index Measuring Device}
%
%
% author names and IEEE memberships
% note positions of commas and nonbreaking spaces ( ~ ) LaTeX will not break
% a structure at a ~ so this keeps an author's name from being broken across
% two lines.
% use \thanks{} to gain access to the first footnote area
% a separate \thanks must be used for each paragraph as LaTeX2e's \thanks
% was not built to handle multiple paragraphs
%

\author{Gianluca Hiss Garbim, Luis Carlos Mathias, André Massami Assakawa and Taufik Abrão% <-this % stops a space
\thanks{G. H. Garbim, L. C. Mathias, and A. M. Assakawa are with the Computer Engineering Coordination, Federal University of Technology - Paran\'a, Toledo, Paran\'a, 85902-490, Brazil (e-mail: mathias@utfpr.edu.br).}% <-this % stops a space
\thanks{T. Abrão is with Departamento de Engenharia Elétrica, Universidade Estadual de Londrina, Londrina-Paran\'a, 86057-970, Brazil}% <-this % stops a space
\thanks{Manuscript received April xx, 2024; revised April xx, 2024.}}

% note the % following the last \IEEEmembership and also \thanks - 
% these prevent an unwanted space from occurring between the last author name
% and the end of the author line. i.e., if you had this:
% 
% \author{....lastname \thanks{...} \thanks{...} }
%                     ^------------^------------^----Do not want these spaces!
%
% a space would be appended to the last name and could cause every name on that
% line to be shifted left slightly. This is one of those "LaTeX things". For
% instance, "\textbf{A} \textbf{B}" will typeset as "A B" not "AB". To get
% "AB" then you have to do: "\textbf{A}\textbf{B}"
% \thanks is no different in this regard, so shield the last } of each \thanks
% that ends a line with a % and do not let a space in before the next \thanks.
% Spaces after \IEEEmembership other than the last one are OK (and needed) as
% you are supposed to have spaces between the names. For what it is worth,
% this is a minor point as most people would not even notice if the said evil
% space somehow managed to creep in.

% The paper headers
%\markboth{IEEE Transactions on Instrumentation \& Measurement,~Vol.~xx, 2026}
%
\markboth{}
{Shell \MakeLowercase{\textit{et al.}}: Bare Demo of IEEEtran.cls for IEEE Journals}

% make the title area
\maketitle

\begin{abstract}
The correlated color temperature (CCT) and color rendering index (CRI) of artificial light sources are important because they have implications for human biology and professional applications.
Although CCT information is generally available for commercial lamps, CRI is commonly not reported.
In addition, devices measuring these parameters are difficult to access as they require a spectrophotometer, a commonly expensive device.
In this context, the present work designs and builds a meter in detail, from the structural part of the equipment, interface with sensors, calculation, to the compensation algorithms implementation, aiming  to build dedicated functionalities of a spectrophotometer, which is designed without the use of optical lenses.
In addition to simplifying the device, this approach allows measurements free from dispersions caused by chromatic aberrations typical of optical lenses.
The prototype obtained proved to be effective, capturing the spectral power distributions of various light sources and calculating their CCT and CRI.
\end{abstract}

% Note that keywords are not normally used for peerreview papers.

% For peer review papers, you can put extra information on the cover
% page as needed:
% \ifCLASSOPTIONpeerreview
% \begin{center} \bfseries EDICS Category: 3-BBND \end{center}
% \fi
%
% For peerreview papers, this IEEEtran command inserts a page break and
% creates the second title. It will be ignored for other modes.
\IEEEpeerreviewmaketitle

% ***************************************************************************8
\section{Introduction}

% Why have quality lighting?
\IEEEPARstart{I}{n} fact, lighting directly affects biological processes, such as the human circadian rhythm \cite{2022_circadian}.
In this context, artificial lighting can harm the quality of rest, professional work, sales, emotions, even affecting the perception of sweet and bitter flavors \cite{tetsuo2005}.

% And why quality in color reproduction?
%\colr{One of the ways} 
A way to evaluate the quality of a light source is by its color temperature, which defines how much warmer or colder the light source is.
However, this parameter alone does not measure the quality of color reproduction in a environment \cite{2004_cri}.
Objects illuminated by a lamp with high color rendering would imply that most people would %\colr{describe their appearance as natural and pleasant} 
describe them as having a natural and pleasant appearance. 
%(see Fig. \ref{fig:cricomp})
On the other hand, if illuminated with an inappropriate lamp, objects would take on a less familiar and distorted appearance.

%\begin{figure}[h]
%   \centering
%   \includegraphics[width=\linewidth]{fig/cricomparacao.jpg}
%   \caption{Objects under lighting with adequate (left side) and inadequate (right side) color reproduction.}
%   \label{fig:cricomp}
%\end{figure}

% Where is CRI most important?
Among the applications in which adequate color reproduction is essential, it is possible to mention any activity that involves the careful choice of colors such as photography, graphic design, lighting, fine arts, as well as those in which the identification of natural processes and organisms, requires attention to color in a controlled environment.
One example is the work of \cite{ito_2022}, which uses neural networks to optimize the light source spectrum for detecting oral lesions in dental applications.

% Measuring instrument
One way to measure this aspect is by the color rendering index (CRI), which is a parameter that consists of comparing the color reproduction of the test light source with a reference source, commonly a model of sunlight.
It is based on mathematical comparisons with the expected color curves, in which it is computed in a single numerical value, with the best possible value being 100~\% and the worst ones being able to assume negative values \cite{cherng2012}.
Although it possesses limitations such as the discontinuity in the color temperature of 5000~K and the reduced number of samples \cite{sandor2006}, it satisfactorily fulfills the role of measuring the naturalness of the color reproduction of an artificial light source \cite{ royer2016}.
For example, this parameter is even used in the design of visible light communication systems, as in addition to transmitting data, the luminaire must be able to illuminate \cite{2023_vlc_cri}.

% Justification
However, in many countries, deficiencies (and even the absence) of regulations and legislation related to lighting quality result in great difficulty in predicting the quality of a luminaire. In most cases, this information does not exist for commercial lamps. Added to this fact, CRI measuring instruments are generally not accessible as they require, for their proper functioning, a spectrophotometer \cite{2017_how_design}.

% Related works
In the literature there are several works that approach the subject of the construction of spectrophotometers \cite{2017_how_design,2024_design_spectrometer}. 
However, most of them are focused on physical/chemical analysis applications and not on determining the quality of artificial lighting.
In these approaches, dispersions in the measurements may occur due to the characteristic of chromatic aberration in optical lenses. This occurs due to the typical variation of the refractive index of the lens material as a function of the wavelength \cite{2021horn}.
The work in \cite{2022_cnn_quality} uses artificial neural networks, image cameras and color samples on printed paper. However it only identifies the light source among those tested, not measuring the color temperature or CRI. 
To date, no work has been found that fully presents the construction of this type of meter.

%\colr{Mathias, confirme a correção, acurácia e completude com reduzida redundância/repetição para os próximos 3 paragrafos.}

In the following, we explicitly address the limitations of existing correlated color temperature (CCT) and CRI measurement devices and explain how the proposed solution overcomes such challenges.

\vspace{1mm}
\noindent\textbf{Problem Statement:}
Despite the critical role that CCT and CRI play in assessing the quality of artificial lighting, existing measurement devices face significant limitations. Firstly, while CCT information is often available for commercial lamps, CRI is frequently not reported, leaving a gap in understanding the color rendering capabilities of these light sources. Secondly, the devices currently available for measuring CCT and CRI, typically require a spectrophotometer, which is an expensive and complex piece of equipment. This inaccessibility restricts the ability of professionals and consumers to evaluate lighting quality effectively.
Moreover, many existing spectrophotometers rely on optical lenses, which can introduce chromatic aberrations and distortions in measurements due to variations in the refractive index of the lens material across different wavelengths. This can lead to inaccuracies in the reported CCT and CRI values, further complicating the selection of appropriate lighting solutions.

\vspace{2mm}
\noindent\textbf{Proposed Solution:}
To address these challenges, the proposed work presents a novel lighting quality meter designed to measure CCT and CRI without the use of optical lenses. This innovative approach simplifies the construction of the device, making it more cost-effective and accessible for users. By employing a dedicated spectrophotometer that utilizes alternative optical arrangements, the device minimizes the risk of chromatic aberrations, thereby enhancing measurement accuracy. This cost-effective solution ensures that professionals and consumers can evaluate lighting quality without the financial burden of expensive equipment.
The comprehensive design integrates both hardware and software components, allowing the effective capture of spectral power distributions (SPDs) from various luminaires. The meter's calibration process, validated against a commercial spectrophotometer, ensures reliable calculations of CCT and CRI. Ultimately, this solution not only provides a practical tool for measuring lighting quality but also empowers users to make informed decisions regarding artificial light sources, thereby improving their applications in fields such as photography, design, and healthcare.

Thus, this work proposes to comprehensively integrate a software solution and a hardware device composed of a dedicated spectrophotometer for analyzing the light spectrum of luminaires, allowing the determination of the CCT and CRI using accessible materials and equipment. In this sense, this research work has the following goals:
\begin{enumerate}
\setlength{\itemsep}{0pt}
     \item prepare a comprehensive review of the quality of artificial lighting, specifically concerning color reproduction;
     \item establish two optical geometric arrangements for the construction of the meter, without the use of optical lenses; therefore, of low-cost, simplified construction and without chromatic aberration;
     \item determination of construction parameters, analysis of expected accuracy and linearity of measurements;
     \item build the meter, from the structural mechanical part, electronic interfaces, calibration and compensation, operating algorithms and calculations of color reproduction quality parameters;
     \item carry out measurements of different luminaires to validate the proposed meter.
\end{enumerate}

The remainder of this document is organized into five sections. Section \ref{sec:reprod_cor} presents the theory of perception and measurement of color reproduction, while Section \ref{sec:geom_opt} 
analyzes the optical geometry of the meter.
Section \ref{sec:construction} presents details of the building and calibration of the meter prototype, while its results from measuring the lighting quality of different luminaires are presented in Section \ref{sec:results}. Finally, the Section \ref{sec:concl} presents the main conclusions of the work.

% *********************************************
\section{Color reproduction} \label{sec:reprod_cor}

\begin{figure}[t]%[htbp]
   \centering   {\includegraphics[width=.42\textwidth]{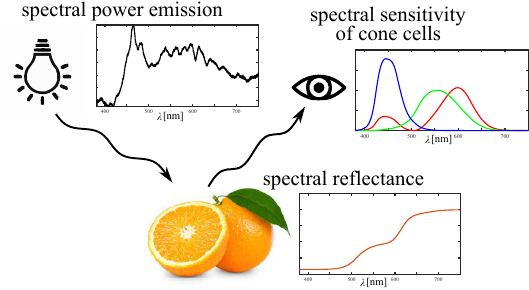}}
   \caption{Color perception process \cite{huete2004,stockman1993}.}
   \label{fig:process}
\end{figure}

The human eye has the ability to perceive colors by capturing light, which is commonly composed of different intensities at different wavelengths.
In this sense, the spectral power distribution (SPD) is the curve that describes the amount of energy that a light source emits as a function of wavelength.
According to Fig. \ref{fig:process}, the light perception process can be divided into three stages. Initially, the illuminant with a specific SPD illuminates an object, that has a specific reflectance spectrum. 
Under suitable lighting, the light reflected by the object is captured by the three types of vision cone cells, each of which has sensitivity in different wavelength ranges \cite{jacobs1996,stockman1993}.
%According to Fig. \ref{fig:lms_xyz}(a), 
%These curves can be represented by $\bar l(\lambda)$, $\bar m(\lambda)$ and $\bar s(\lambda)$, for lengths long, medium and short wave (LMS) \cite{stockman1993}.
%These operate effectively in the photopic region, under suitable lighting.
%\footnote{In the scotopic region, with low illumination, rod cells predominate, which present less distinction in terms of colors\cite{pericles1990}. Therefore, in the analysis of color perception, they are not considered.}

\begin{figure}[t]
   \centering
    %\subfloat[]{\includegraphics[width=.4\textwidth]{fig/lms.eps}} \\
    %\subfloat[]
    {\includegraphics[width=.42\textwidth]{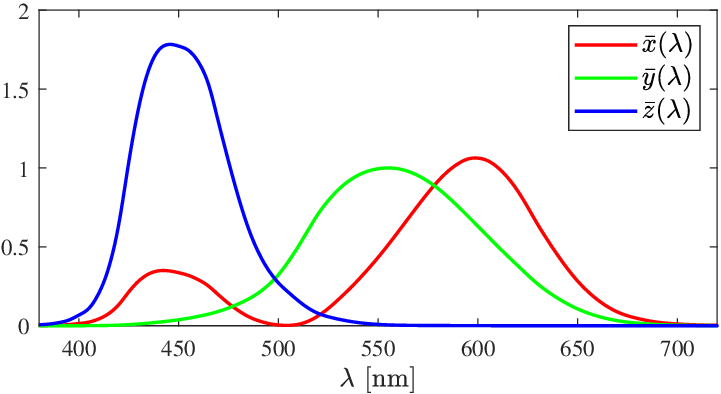}}
   \caption{
   %Relative sensitivity of cone cells \cite{stockman1993} in (a) and 
   Color matching functions.}
   \label{fig:lms_xyz}
\end{figure}

%--------------------------------------------------
\subsection{Color spaces} \label{subsec:color_spaces} 

Three values, called tristimulus, were developed to represent the stimulus level of each type of cone receptor.
In the 1931 International Commission on Illumination (CIE) color model, the tristimulus $X$, $Y$, and $Z$ are used \cite{smith1931}. $Y$ represents luminance, while $X$ and $Z$ represent chromaticity coordinates. The great advantage of this scheme is that, for any luminance value, a plane can represent all possible chromaticities.

Due to the physical distribution of cones in the eye, tristimulus values are affected by the observer's field of vision \cite{britannica}. To mitigate the influence of this variable, the CIE 1931 establishes a colorimetric standard observer, with the purpose of standardizing an observer's chromatic response to an arc of 2$^{\circ}$ from the fovea (ocular region where most are concentrated cone cells).
To numerically describe this chromatic response, color matching functions $\bar{x}$, $\bar{y}$ and $\bar{z}$ were developed, which are available in the table 2DegStdColObs (2 degree standard color observer) in \cite{smith1931}.
These are demonstrated in Fig. \ref{fig:lms_xyz}.
%and are used instead of the sensitivity functions of the individual cones ($\bar l(\lambda)$, $\bar m(\lambda)$ and $\bar s(\lambda)$) because it is closer to the way the standard observer responds to stimulus of each color.

In this way, the tristimulus values $X$, $Y$ and $Z$ can be defined by product integration between the SPD $S(\lambda)$ of the captured light and the color matching functions:
\begin{equation}
   X=\int _{\lambda _{\textsc{l}}}^{\lambda_{\textsc{h}}}S\!\left(\lambda \right)\bar{x}\!\left(\lambda \right)d\lambda;
   \label{eq:X}
\end{equation}
\begin{equation}
   Y=\int _{\lambda _{\textsc{l}}}^{\lambda_{\textsc{h}}}S\!\left(\lambda \right)\bar{y}\!\left(\lambda \right)d\lambda;
   \label{eq:Y}
\end{equation}
\begin{equation}
   Z=\int _{\lambda _{\textsc{l}}}^{\lambda_{\textsc{h}}}S\!\left(\lambda \right)\bar{z}\!\left(\lambda \right)d\lambda;
   \label{eq:Z}
\end{equation}
where $\lambda_{\textsc{l}}$ represents the shortest wavelength for which the SPD presents information, and $\lambda_{\textsc{h}}$ the longest \cite{smith1931}. In this work, the variables that have the subscript $_\textsc{l}$ and $_\textsc{h}$ are relative to the lower and upper wavelength limits, respectively.

From $X$, $Y$ and $Z$, it is possible to determine the parameters $x$, $y$ and $z$ that define the chromaticity in the color space:
\begin{equation}
   x=\frac{X}{X+Y+Z};
   \label{eq:x}
\end{equation}
\begin{equation}
   y=\frac{Y}{X+Y+Z};
   \label{eq:y}
\end{equation}
\begin{equation}
   z=\frac{Z}{X+Y+Z}=1-x-y.
   \label{eq:z}
\end{equation}
where the value of $z$ can be deduced from $x$ and $y$, it is generally discarded in the representations, giving rise to the CIE 1931 color space.
%, as shown in Fig. \ref{fig:results}(a). % \ref{fig:cie1931}. 
%This color space is also called the CIE xyY color space \colb{and}, the luminance $Y$ is used as the third descriptive parameter \cite{smith1931}.

%\begin{figure}[t]
%   \centering
%   \includegraphics[width=.37\textwidth]{fig/ciexy1931.pdf}   
%   \caption{CIE 1931 space color \cite{smith1931}. At the edges, the blue values refer to single wavelength light.}
%   \label{fig:cie1931}
%\end{figure}

A notable difficulty of the CIE 1931 color space is that it is not uniform, that is, the Euclidean distances between points on the graph are not proportional to differences in the observer's visual perception.
As a solution, the uniform color space (UCS) was developed, which presents chromatic distances proportional to those perceived by the human eye in coordinates $u$ and $v$ \cite{judd1935}.
This was simplified in 1937 and became the standard recommended by the CIE in 1960 \cite{macadam1937}. 
%The results in Fig. \ref{fig:results}(b) are an example of using this color space. 
This can be obtained by transforming chromaticities:
\begin{equation}
   u=\frac{4x}{12y-2x+3};
   \label{eq:u1}
\end{equation}
\begin{equation}
   v=\frac{6y}{12y-2x+3}.
   \label{eq:v1}
\end{equation}

Later, the CIE 1964 color space 
%(or CIE UVW) 
was developed to enable the calculation of color differences without the need to maintain constant luminance $Y$. This uses a luminosity index $W^\star$ together with the chromaticity coordinates $U^\star$ and $V^\star$ \cite{macadam1937}.
For the purpose of lighting quality analysis, it is possible to transform colorimetric data to the CIE 1964 color space through the following relationships:
\begin{equation}
   W^{\star}=25Y^{\frac{1}{3}}-17;
   \label{eq:Wstar}
\end{equation}
\begin{equation}
   U^{\star}=13W^{\star}\left(u_k-u_r\right);
   \label{eq:Ustar}
\end{equation}
\begin{equation}
   V^{\star}=13W^{\star}\left(v_k-v_r\right);
   \label{eq:Vstar}
\end{equation}
in which $u_r$ and $v_r$ represent the chromaticity of the reference, $u_k$ and $v_k$ represent the chromaticity of the light source being analyzed, all from the CIE 1960 color space. With the reference and illuminant data transformed into this color space, color differences can be calculated.

%--------------------------------------------------
\subsection{Correlated color temperature}\label{sec:cct}

A blackbody is a theoretical object that absorbs all incident electromagnetic radiation, emitting radiation \cite{planck1914}.
Taking into account that LED and fluorescent lamps do not have the same SPD as black bodies, the correlated color temperature (CCT) was developed to relate the light source to the nearest black body. However, contrary to the conventional temperature scale, the color temperature scale goes from warm (lower temperatures) to cool (higher temperatures) \cite{ashdown2002}.
For example, the color temperature of sunlight varies from warmer values at sunrise and sunset (about 3000~K) to cooler values near midday (in the range of 5000 and 6000~K).

Two methods to approximate the CCT are described in \cite{hernandez1999}. The first uses a polynomial equation:
\begin{equation}
   T_{\textsc{cct}}(x,y)=-449{n}^3+3525n^2-6823.3n+5520.33;
   \label{eq:ccta}
\end{equation}
in which $n = \frac{x-x_e}{y-y_e}$ with epicenter at $x_e=0.3320$ and $y_e=0.1858$ \cite{maccamy1992}. 
However, this presents a large error for very high CCTs \cite{hernandez1999}.
The second uses the following
sum of exponentials equation:
\begin{equation}
\resizebox{.98\hsize}{!}{$
T_{\textsc{cct}}\left(x,y\right)={A_0}+{A_1}\exp\left(-\frac{n}{{t_1}}\right)+{A_2}\exp\left(-\frac{n}{{t_2}}\right)+{A_3}\exp\left(-\frac{n}{{t_3}}\right); $}
   \label{eq:cctb}
\end{equation}
considering the epicenter at $x_e=0.3366$ and $y_e=0.1735$; and the coefficients, for lighting purposes, between 3000 and 50000~K, 
are $A_0=-949.8631$, $A_1= 6253.80338$, $A_2=28.70599$, $t_1=0.92159$, $t_2=0.20039$ and $t_3=0.00004$.
%given in Tab. \ref{tab:coef}.

% \begin{table}[!hbt]
% %\captionsetup{width=\textwidth}
% \centering
% \caption{ Coefficients for sunlight  \cite{hernandez1999}.}
% \label{tab:coef}
% \begin{tabular}{cc|cc}
% 	\hline
% 	\textbf{Parameter}  & \textbf{Value} & \textbf{Parameter}  & \textbf{Value}  \\ \hline
%    $A_0$               & -949.86315     & $t_1$               & 0.92159 \\
%    $A_1$               & 6253.80338     & $t_2$               & 0.20039 \\
%    $A_2$               & 28.70599       & $t_3$               & 0.07125 \\
%    $A_3$               & 0.00004        & \\ \hline
% \end{tabular}
% \end{table}

%--------------------------------------------------
\subsection{Standard illuminant}\label{subsec:spd}

The daylight is chosen as the reference because it is a lighting condition commonly perceived as natural and pleasant, while blackbody radiators are chosen for lower temperatures because they present similar behavior to incandescent and halogen lamps.
%In this sense, the reference lighting source must be a reconstruction of daylight for CCTs greater than or equal to 5000~K (standard illuminant D). For smaller CCTs, a blackbody radiator (standard A illuminant) must be used \colb{\cite{schanda2007}}.
The spectral energy distribution of a blackbody can be approximated as a function of its CCT.
Therefore, Planck's radiation law is a suitable approximation for the SPD of a lamp with a color temperature below 5000~K (standard A illuminant) \cite{planck1914,schanda2007}:
\begin{equation}
   S_{\textsc{a}}(\lambda)=\frac{c_1}{\lambda^5} \cdot \left[\exp\left(\frac{c_2}{\lambda}\cdot T_{\textsc{cct}}\right)-1\right]^{-1} \qquad \left[ \frac{W}{m^3} \right];
   \label{eq:refcorponegro}
\end{equation}
where $c_1 = 3.7418\times10^{-16}$ and $c_2 = 1.4388 \times 10^{-2}$ are the Planck radiation constants.

For CCTs above 5000~K, it is possible to approximate the SPD of the reference source using the D series illuminant through the composition of three SPDs, $S_0$, $S_1$ and $S_2$, available in the CIEDaySN table in \cite{smith1931}, which were obtained from empirical measurements. $S_0$ corresponds to the average of all SPD sunlight samples; $S_1$ to yellow-blue variations due to the presence of clouds; and $S_2$ to pink-green variations due to the presence of water vapor.
In this approach, the SPD is:
\begin{equation}
   S_{\textsc{d}}(\lambda)=S_0(\lambda)+M_1 S_1(\lambda)+M_2 S_2(\lambda)
   \label{eq:spdaprox};
\end{equation}
in which the coefficients $M_1$ and $M_2$ are determined by:
\begin{equation}
   M_1=\frac{-1.3515-1,7703x_\textsc{d}+5.9114y_\textsc{d}}{0.0241+0.2562x_\textsc{d}-0.7341y_\textsc{d}};
   \label{eq:m1}
\end{equation}
\begin{equation}
   M_2=\frac{0.030-31,4424x_\textsc{d}+30.0717y_\textsc{d}}{0.0241+0.2562x_\textsc{d}-0.7341y_\textsc{d}};
   \label{eq:m2}
\end{equation}
where the values of $x_\textsc{d}$ and $y_\textsc{d}$ originate from the chromaticities defined by the CIE 1931 color space, which are determined by their correlated color temperature $T_{\textsc{cct}}$:
%for the illuminant D:
\begin{equation}
   y_\textsc{d} =-3x_\textsc{d}^2+2.87x_\textsc{d}-0.275.
   \label{eq:yd}
\end{equation}
For 4000~K~$\le T_{\textsc{cct}}\le$~7000~K we have:
\begin{equation}
   x_\textsc{d}= 0.244063+0.09911\frac{10^3}{T_{\textsc{cct}}}+2.9678 \frac{10^6}{T_{\textsc{cct}}^2}-4.6070\frac{10^9}{T_{\textsc{cct}}^3};
   \label{eq:xd1}
\end{equation}
already for 7000~K~$\le T_{\textsc{cct}}\le$~25000~K:
\begin{equation}
   x_\textsc{d}= 0.237040+0.24748\frac{10^3}{T_{\textsc{cct}}}+1.9018\frac{10^6}{T_{\textsc{cct}}^2}-2.0064\frac{10^9}{T_{\textsc{cct}}^3}.
   \label{eq:xd2}
\end{equation}

\begin{figure}[t]
   \centering
   {\includegraphics[width=.42\textwidth]{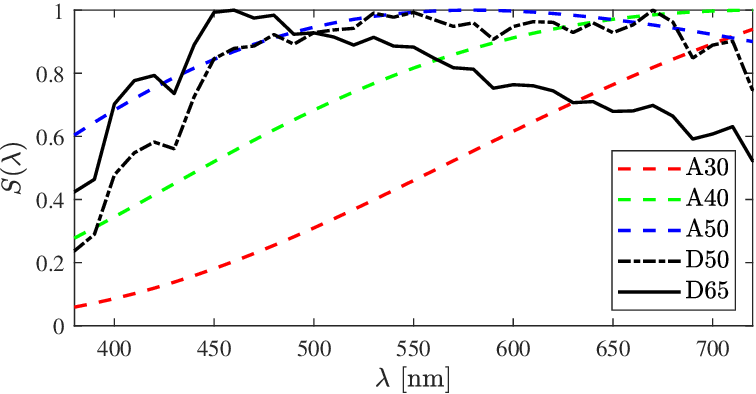}}
   \caption{Normalized SPD of standard type A and D illuminants.}
   \label{fig:spd_ref}
\end{figure}

Fig. \ref{fig:spd_ref} depicts the SPDs, normalized by the peak, for standard illuminants A for CCTs of 3000, 4000 and 5000~K (A30, A40 and A50) and also for standard illuminants D for CCTs of 5000 and 6504~K (D50 and D65). As expected, and in general, the increase in CCT implies an increase in intensity at shorter wavelengths compared to longer wavelengths. Another detail is that for the same CCT of 5000~K, the D50 illuminant appears to follow the behavior of the A50 illuminant curve from approximately 450 nm onwards.  

%--------------------------------------------------
\subsection{Metamerism and chromatic adaptation} \label{subsec:metam}

%There are two phenomena that make the quantification of color perception more complex: the metamerism of light sources and the chromatic adaptation that human vision performs to preserve the relative appearance of colors.

It is possible for light sources with different SPDs to present equal chromaticities. This behavior is called metamerism. For example, a blackbody whose SPD is represented by a curve can generate the same chromaticity as an LED light source with several peaks spread across the spectrum. However, the perceived color reproduction will still be different \cite{setchell2012}.

Another phenomenon that affects color perception is chromatic adaptation. Process carried out by the human visual system that adapts the perception of a color to match what is expected. For example, a painting observed under midday sunlight has very different colors than the same one observed under sunset lighting. Our visual system, however, adapts these colors so that they maintain their relative differences regardless of these variations \cite{luo2014}.

Numerically, chromatic adaptation can be carried out using the $u$ and $v$ parameters of the CIE 1960 color space \cite{cie1995}:
\begin{equation}
   c=\frac{4-u-10v}{v};
   \label{eq:CA-c}
\end{equation}
\begin{equation}
   d=\frac{1.708v+0.404-1.481u}{v}.
   \label{eq:CA-d}
\end{equation}

These are calculated both for the light source under evaluation, $c_k$ and $d_k$, and for the reference samples, $c_r$ and $d_r$. Thus, the new chromaticity coordinates for a given sample $i$ can be defined by:
\begin{equation}
   u'_{k,i}=\frac{10.872+0.404\frac{c_r}{c_k}c_{k,i}-4\frac{d_r}{d_k}d_{k,i}}{16.518+1.481\frac{c_r}{c_k}c_{k,i}-\frac{d_r}{d_k}d_{k,i}};
   \label{eq:CA-u'ki}
\end{equation}
\begin{equation}
   v'_{k,i}=\frac{5.520}{16.518+1.481\frac{c_r}{c_k}c_{k,i}-\frac{d_r}{d_k}d_{k,i}};
   \label{eq:CA-v'ki}
\end{equation}
in which $u'_{k,i}$ and $v'_{k,i}$ represent the adapted colors and $d_{k,i}$ refers to the product of the SPD of the light evaluated light source and the reflectance of sample $i$.

%--------------------------------------------------
\subsection{Color rendering index} \label{sec:cri}

The color rendering index (CRI) evaluates, through comparison with a reference, the quality of color reproduction of artificial light sources. % \cite{cie1995}. 
Conventionally, the appearance of predefined color samples under the effect of a standard illuminant is compared with the reproduction of the illuminant being tested.
The differences are described by the distance between the sample illuminated by the reference lamp and the lamp under evaluation in terms of chromaticity and luminance \cite{cie1995}.

This way, it is possible to calculate the CRI for specific samples, which are described by a specific reflectance function.
For each sample, the distance between the color generated by the test source and the color generated by the reference (both in the CIE 1964 color space) is calculated \cite{cie1995}:
\begin{equation}
\resizebox{.98\hsize}{!}{$
   \Delta E_i \!=\!\sqrt{\left(\Delta U^{\ast }_{r,i}\!-\!\Delta \:U^{\ast \:}_{k,i}\right)^2\!+\!\left(\Delta V^{\ast \:}_{r,i}\!-\!\Delta V^{\ast \:\:}_{k,i}\right)^2\!+\!\left(\Delta W^{\ast \:\:}_{r,i}\!-\!\Delta W^{\ast \:\:\:}_{k,i}\right)^2}\!; $}
   \label{eq:deltaE}
\end{equation}
then generating the CRI value for this sample:
\begin{equation}
   R_i=100-4.6\Delta E_i.
   \label{eq:Ri}
\end{equation}

\begin{figure}[t]
   \centering
   {\includegraphics[width=.42\textwidth]{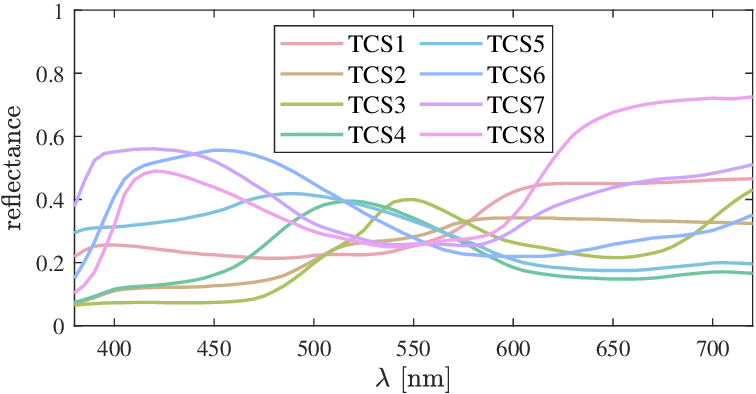}}
   \caption{Reflectances of test color samples \cite{smith1931}.}
   \label{fig:reflectance_samples}
\end{figure}

To calculate the average CRI $\mathrm{R}_a$ of a light source, eight test samples (TCS) must be illuminated by the source under evaluation and, subsequently, by the reference source \cite{cie1995}.
This process can occur in a simulated way, applying the SPD measured directly from the light source to be tested and the SPD from the reference source, both on the reflectance functions of each sample (see Fig. \ref{fig:reflectance_samples}). Thus, the general value of the CRI can be obtained from the arithmetic mean:
\begin{equation}
   \mathrm{R}_a =\frac{1}{8}\sum _{i=1}^8\:{R}_i.
   \label{eq:CRI}
\end{equation}

It should be noted that the CRI $\mathrm{R}_a$ has a maximum value of 100~\%, in which the lamp reproduction is virtually identical to that of the chosen reference.
At CRIs below this value, the difference to 100 is proportional to the average magnitude of the color differences.
In practice, differences of the order of 5 are not noticeable to an observer \cite{cie1995}.

% *********************************************
\section{Meter Optical geometry} \label{sec:geom_opt}

SPD can be measured using a device called a spectrophotometer. In general, this can be built using a light scatterer, lenses, and photosensors \cite{2017_how_design,2024_design_spectrometer}.
As a disperser, diffraction gratings are commonly used, which are devices that have hundreds to thousands of slits per millimeter.
As photosensors, a linear CCD (charge-coupled device) sensor can be used, a device that has thousands of photosensors arranged in a line, which are generally used in barcode reader devices \cite{ccd}.

As in this work, the light intensities generated by the luminaires are much greater than those commonly used in chemical analysis spectrophotometers, the project can become simplified and more compact without the necessity of optical lenses. This approach has the advantage that measurements do not suffer from chromatic aberrations, which are common when optical lenses are used.

\begin{figure}[t]
   \centering
    \subfloat[]{\includegraphics[width=.36\textwidth]{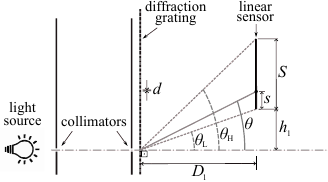}} \\
    \subfloat[]{\includegraphics[width=.36\textwidth]{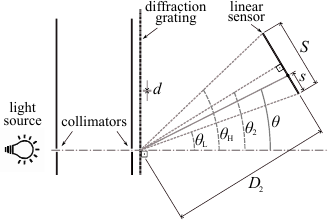}}
   \caption{Geometry of the first arrangement in (a) and the second arrangement in (b).}
   \label{fig:arranjo}
\end{figure}

As shown in Fig. \ref{fig:arranjo}(a), after the collimating bulkheads, the light passes through the diffraction grating in several slits, and in each slit it is diffracted at different angles. 
This causes the light rays of the same wavelength, which reach a point on the sensor $s$, to travel different distances and thus reach a point $s$ with different time delays.
Considering the wave characteristic of light, destructive or constructive interference may occur if the delays are close to odd multiples or even multiples of half a wavelength, respectively \cite{halliday2021}.
Thus, an interference figure is created along the linear CCD sensor, where the position of the maximum point $s$, where constructive interference occurs, depends on the wavelength of the light that falls on the diffraction grating \cite{halliday2021}:
\begin{equation}
    d \sin\theta = m \lambda;
    \label{eq:maximo_interf}
\end{equation}
in which $d$ is the spacing between the slits, $\theta$ the angle between the central axis and the diffracted rays, $m$ the order of the position being measured, and $\lambda$ the wavelength being evaluated.

In this way, each wavelength will have a different maximum point on the sensor, allowing the intensities to be discriminated depending on the wavelength, \textit{i.e.}, the SPD.
In this sense, this section models and adequately designs the arrangement of the components of a spectrophotometer, considering two distinct geometric arrangements.
In both analyses, the objective is to use the entire detector length of the linear CCD sensor in order to obtain a greater number of detectors per wavelength range analyzed. This directly implies a greater resolution of the meter, increasing the number of points of the SPD curve captured.

%--------------------------------------------------
\subsection{CCD sensor parallel to the diffraction grating}

This first approach considers a classic arrangement that is very common in the literature \cite{halliday2021}.
The angles $\theta_{\textsc{l}}$ and $\theta_{\textsc{h}}$ can be determined by 
\eqref{eq:maximo_interf} for $\lambda_{\textsc{l}}$ and $\lambda_{\textsc{h}}$. 
The first order ($m = 1$) will be considered, as it has the advantage of presenting greater angular dispersion and, consequently, 
enabling a smaller distance between the diffraction grating and the linear CCD sensor given the total range of $\lambda$ to be captured.
%a better relationship  between the captured $\lambda$ resolution and the distance between the diffraction grating and the linear CCD sensor.

From the smaller right triangle in Fig. \ref{fig:arranjo}(a), it is possible to determine the height $h_1$:
\begin{equation}
    h_1 = D_1 \tan \theta_{\textsc{l}}
    \label{eq:theta_L1}
\end{equation}
and from the larger right triangle we have:
\begin{equation}
    S + h_1 = D_1 \tan \theta_{\textsc{h}};
    \label{eq:theta_H1}
\end{equation}
where $S$ is the useful length of the linear CCD sensor.

By applying \eqref{eq:theta_L1} to \eqref{eq:theta_H1}, it is possible to determine the distance:
\begin{equation}
    D_1 = \frac{S}{\tan \theta_{\textsc{h}} - \tan \theta_{\textsc{l}}};
\end{equation}
finally allowing us to determine the height $h_1$ from \eqref{eq:theta_L1}.

%--------------------------------------------------
\subsection{Linear CCD sensor inclined to the diffraction grating}

In this second arrangement, the linear CCD sensor is positioned perpendicular to a medium angle:
\begin{equation}
    \theta_{2} = \frac{\theta_{\textsc{h}}+\theta_{\textsc{l}}}{2}.
    \label{eq:angulo_medio}
\end{equation}

According to Fig. \ref{fig:arranjo}(b), the sensor forms the base of an isosceles triangle. Through the upper right triangle, it is possible to obtain the last parameter of the project:
\begin{equation}
    D_2 = \frac{S}{2\tan(\theta_{\textsc{h}}-\theta_{2})}.
    \label{eq:d2}
\end{equation}

Thus, it appears that the calculations for the design of this second arrangement are more expeditious, compared to the previous one.

%--------------------------------------------------
\subsection{Linearity analysis} \label{subsec:lin}

Nonlinearities directly imply distortion of the captured SPD, propagating an error in the calculation of the CCT and CRI. In this sense, it is possible to evaluate the linearity between the position $s$ of the photosensors of the linear CCD sensor and the wavelength $\lambda$ to be detected. Thus, from the analysis of Fig. \ref{fig:arranjo}(a), \eqref{eq:theta_H1} and \eqref{eq:maximo_interf}, it is possible to obtain for the first arrangement:
\begin{equation}
    s_1(\lambda) = D_1 \tan \left( \sin^{-1}\frac{\lambda}{d} \right) -h_1;
    \label{eq:lin1}
\end{equation}
where $\lambda$ is the analyzed wavelength.

Using Fig. \ref{fig:arranjo}(b) and \eqref{eq:d2}, a similar analysis can be done for the second arrangement:
\begin{equation}
    s_2(\lambda) = \frac{S}{2}-D_2\tan \left( \theta_{2}-\sin^{-1}\frac{\lambda}{d} \right).
    \label{eq:lin2}
\end{equation}

If it is exactly linear, the slope must be constant throughout $\lambda$. Therefore, the derivative as a function of wavelength can be applied to \eqref{eq:lin1} and \eqref{eq:lin2}:
\begin{equation}
    \frac{ds_1}{d\lambda} = \frac{dD_1}{(d^2-\lambda^2) \sqrt{1-\frac{\lambda^2}{d^2}}}
    \label{eq:lin_deriv_1}
\end{equation}
and
\begin{equation}
    \frac{ds_2}{d\lambda} = \frac{D_2 \sec^2 (\theta_{2}-\sin^{-1}\frac{\lambda}{d})}{d \sqrt{1-\frac{\lambda^2}{d^2}}}.
    \label{eq:lin_deriv_2}
\end{equation}

\begin{figure}[t!]
   \centering
    \subfloat[]{\includegraphics[width=.42\textwidth]{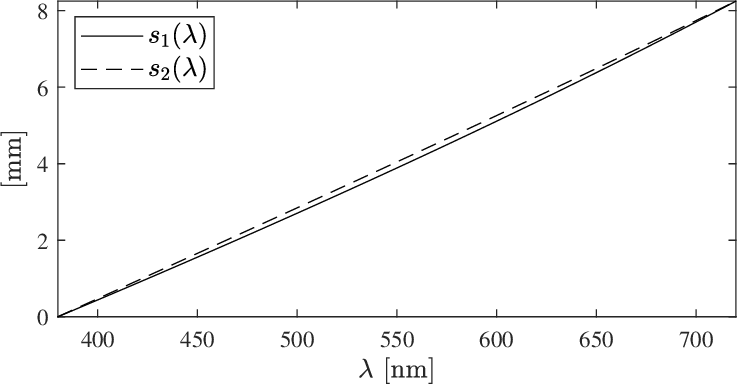}} \\
    \subfloat[]{\includegraphics[width=.42\textwidth]{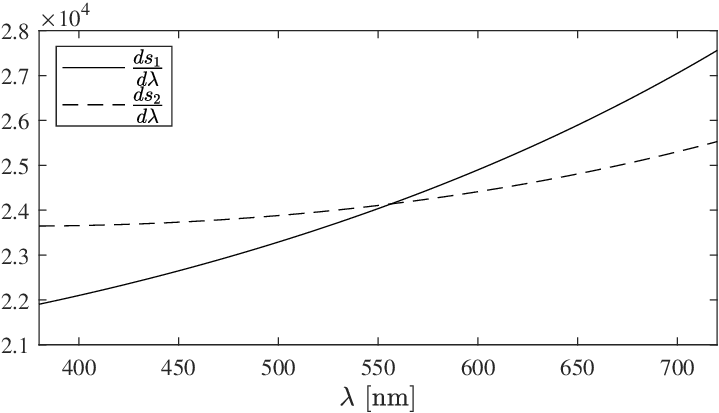}}
   \caption{Position of the maximum point in (a) and its derivative in (b), both as a function of lambda.}
   \label{fig:lin}
\end{figure}

Therefore, neither arrangement is exactly linear. However, as the values of $\lambda$ are much smaller compared to the values of $d$, it is possible to verify that the variation of \eqref{eq:lin_deriv_1} and \eqref{eq:lin_deriv_2} as a function of $\lambda$ are small. Thus, considering the project of this work, for a wavelength range between 380 and 720~nm, a diffraction grating with 600 slits per mm, \textit{i.e.}, $d=1.666$~\textmu m, and a CCD sensor length of $S=8.25$~mm \cite{ccd}, Fig. \ref{fig:lin}(a) was generated, of the maximum position $s$ as a function of $\lambda$. It shows that the second arrangement has better linearity compared to the first arrangement. This is corroborated by Fig. \ref{fig:lin}(b), of its derivatives, which confirms a smaller variation in the slope rate for the second arrangement. 
Therefore, as it presents better linearity characteristics, this arrangement was chosen for the development of this work.

% -----------------------------------
\subsection{Resolution as a function of collimator aperture}

\begin{figure}[t]
   \centering
   {\includegraphics[width=.37\textwidth]{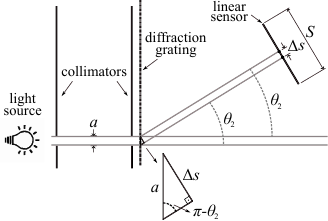}}
   \caption{Resolution as a function of collimator aperture.}
   \label{fig:slit}
\end{figure}

The resolution of the interference figure projected on the linear sensor is dependent on the aperture $a$ of the collimator that allows light to enter the diffraction grating \cite{halliday2021}. This analysis can be best exemplified in Fig. \ref{fig:slit}. The useful length $S$ of the linear sensor must capture the entire range from $\lambda_{\textsc{l}}$ to $\lambda_{\textsc{h}}$. Considering $\delta$ as the required wavelength resolution of the spectrophotometer and that the variation of the position on the sensor $\Delta s$ is approximately linear in relation to the variation of the light wavelength, we have: 
\begin{equation}
    \Delta s \simeq \frac{\delta S}{\lambda_{\textsc{h}}-\lambda_{\textsc{l}}}.
\end{equation}
Thus, based on the right-angled triangle in Fig. \ref{fig:slit}, it is possible to establish the following relationship for determining the aperture:
\begin{equation}
    a \simeq \frac{\delta S}{(\lambda_{\textsc{h}}-\lambda_{\textsc{l}})\sin(\pi-\theta_2)}.
    \label{eq:a}
\end{equation}

Thus, a smaller aperture implies a higher resolution in the measurement; however, it also reduces the amount of captured light. This last consequence is not a problem for the application of this work since in tests carried out, the light sources presented sufficient luminous power to excite the linear CCD sensor, even with apertures of a few tenths of a millimeter.

% *********************************************
\section{Construction and implementation}\label{sec:construction}

\begin{figure}[t]
    \centering
    \subfloat[]{\includegraphics[width=.46\textwidth]{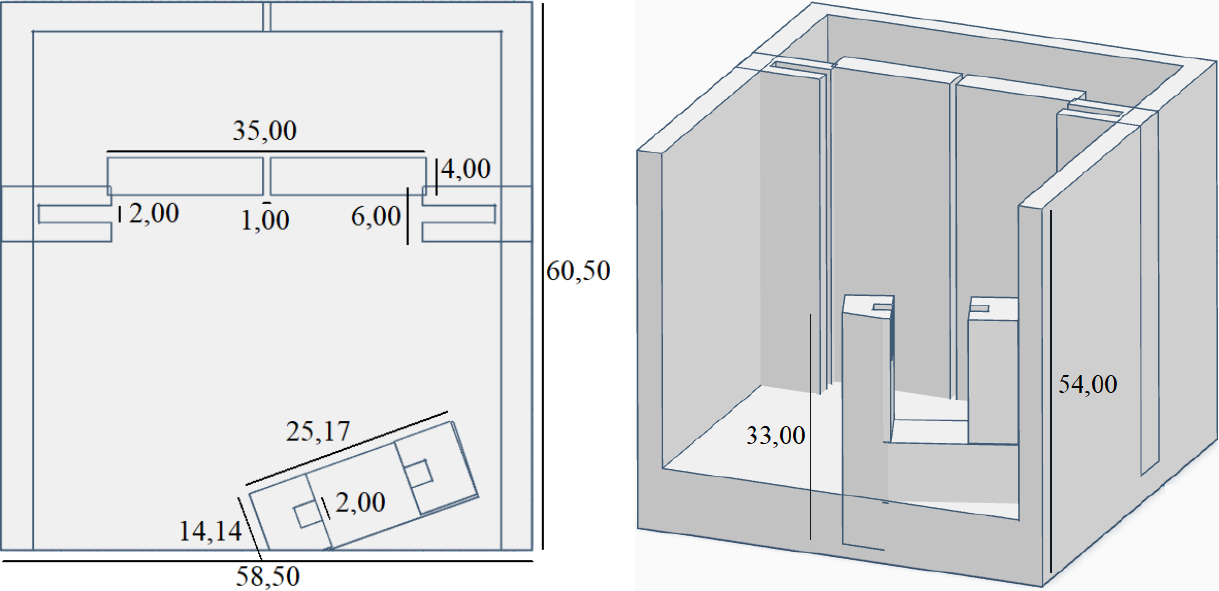}} \\
    \subfloat[]{\includegraphics[width=.4\textwidth]{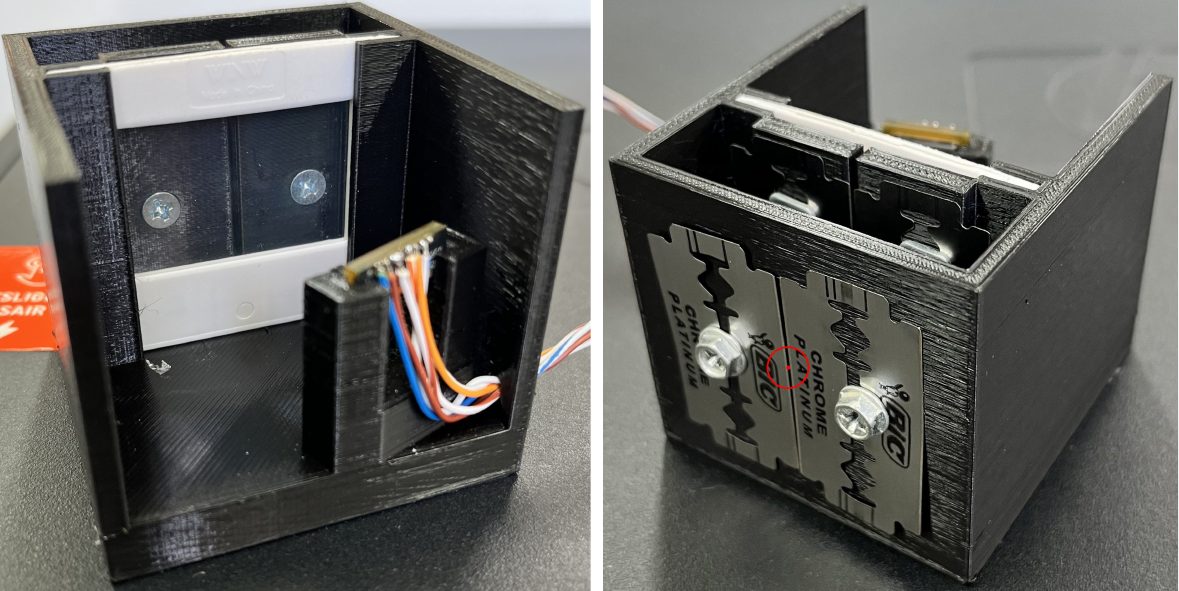}}
    \caption{Geometry and dimensions in mm of the constructed prototype in (a), and internal photos in (b). It is possible to see the installation of the stainless steel blades as collimators, the housings of the diffraction grating, and the CCD linear sensor. The red circle in the photo highlights the opening for light entry.}
    \label{fig:dim_foto}
\end{figure}

% Mechanical structure
Part of the spectrophotometer design parameters were presented in subsection \ref{subsec:lin}, which allows, through \eqref{eq:angulo_medio} and \eqref{eq:d2}, to determine the average angle $\theta_2=33.8 ^\circ$ and the distance $D_2=3.79$~cm, respectively. Therefore, according to Fig. \ref{fig:dim_foto}(a), the design of the structural part of the prototype was carried out using 3-D printing. 
%\lcm{
Fig. \ref{fig:dim_foto}(b) shows the structure printed with Polylactic Acid (PLA) material, highlighting the fitting positions for the diffraction grating, the linear CCD sensor and the two pairs of blades that function as collimators. 
The 3-D printer used was the Creality\textsuperscript{\textregistered} Ender-3, which the manufacturer guarantees accuracy of $\pm$0.1~mm.
%}
The glass protection of the linear CCD sensor was removed to avoid possible reflections, which could distort the measurements.
Considering an expected resolution of $\delta=2.5$~nm, the slot width of the screens was adjusted to $a\simeq0.18$~mm, taking into account \eqref{eq:a}. Also, an ''L''-shaped cover was developed, which can be removable, protecting the device from ambient light in order to only allow, through the collimators, the entry of the light that must be analyzed.  

% -----------------------------------
\subsection{Sensor and microcontroller interface}

\begin{figure}[t]
    \centering
    \subfloat[]{\includegraphics[width=.25\textwidth]{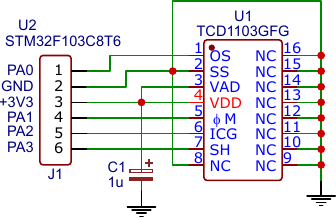}} \\
    \subfloat[]{\includegraphics[width=.47\textwidth]{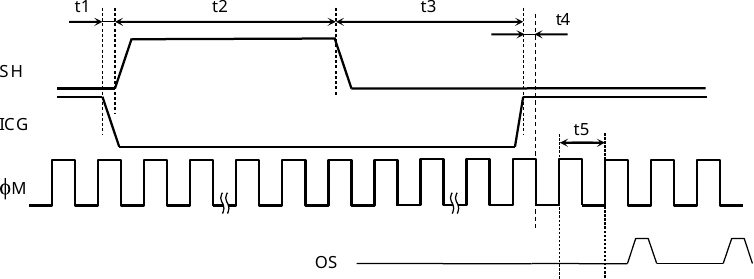}}
    \caption{Electrical diagram in (a) and timing diagram in (b), both of the TCD1103 and \textmu C sensor interface. }
    \label{fig:sch_time}
\end{figure}

According to Fig. \ref{fig:sch_time}(a), a microcontroller (\textmu C) was used as an interface between the TCD1103 sensor and a personal computer (PC).
The OS pin (\textit{output signal}) is an analog output with a voltage proportional to the inverse of the light intensity captured by each sensor. This signal is updated depending on the digital signals ICG (integration clear gate), SH (shift gate), and $\phi$M (master clock) that must be generated by \textmu C.
Three pins are used to power the sensor (SS, $\mathrm{V_{DD}}$ and $\mathrm{V_{AD}}$). The remaining pins are unconnected (NC).
% Frame
After the ICG signal rise sequence (see Fig. \ref{fig:sch_time}(b)), OS presents the reading from the first photosensor. Every two $\phi$M cycles, the TCD1103 switches the reading to the next photosensor.
According to \cite{ccd},
%Fig. \ref{fig:quadro_ccd}, 
the set of photosensors protected from light, signals ranging from D16 to D28, allows the calibration of the dark level, while the effective outputs of the photosensors represent the signals between S1 and S1500.

% Microcontroller and timing constraints
According to Fig. \ref{fig:sch_time}(b), 
%and Table \ref{tab:time_ccd}, 
the timing requirements required by the TCD1103 are very restricted. Therefore, the \textmu C model STM32F103C8T6 from the manufacturer STMicroelectronics was used at the maximum clock frequency of 72~MHz of its ARM Cortex M3 core. A channel of a timer peripheral was configured to generate a pulse width modulated (PWM) signal with a cyclic ratio of 50\%, which corresponds to the rectangular wave of the $\phi$M signal. The interrupt handling function of this timer, associated with an algorithm, allowed the generation of SH and ICG signals.
The implemented times are t1~=~124 t2~=~2500, t3~=~10200, t4~=~7900 and t5~=~2500, all in nanoseconds and within the limits specified in \cite{ccd}.
The analog OS signals were captured using an internal analog-to-digital converter (ADC) channel of \textmu C with 12~bit resolution, which is stored in a vector to later be transmitted to the PC through a serial communication interface. 

%\lcm{
A longer interval between positive SH pulses t$_{\rm int}$ defines a longer capture time in the CCD sensors \cite{ccd}.
Through tests, a t$_{\rm int} \approx 250$~ms allowed adequate capture of the SPD at light intensities starting from 200 lux.
For the lowest power lamp analyzed, the 15~W incandescent, this was equivalent to a distance of approximately 30~cm from the prototype apperture.
%}

% -----------------------------------
\subsection{Responsivity compensation of linear CCD sensors}

% \begin{figure}[t]
%     \centering    {\includegraphics[width=.4\textwidth]{fig/curva_ccd.eps}}
%     \caption{Relative CCD responsivity.}
%     \label{fig:curva_ccd}
% \end{figure}

The linear CCD sensor used in this work is model TCD1103 from the manufacturer TOSHIBA, which has 1500 sensor elements composed of high sensitivity PN photodiodes. Its responsiveness curve is shown in 
%Fig. \ref{fig:curva_ccd} 
\cite{ccd}, as well as its adjustment by a 5th order polynomial function by:
\begin{equation} 
\resizebox{.99\hsize}{!}{$
\begin{array}{l}
     R_{\textsc{ccd}} (\lambda) = -1.783\!\times\!10^{32} \lambda^5 + 4.289\!\times\!10^{26} \lambda^4
     -3,919\!\times\!10^{20} \lambda^3
     \\ \qquad \qquad \quad
     + 1.575\!\times\!10^{14}\lambda^2 -2.170\!\times\!10^{7}\lambda + 0.2012.
     \end{array} $}
     \label{eq:curva_ccd}
\end{equation}
%Therefore, it is necessary to compensate this curve to minimize distortions in the captured intensities as a function of $\lambda$.

% -----------------------------------
\subsection{Calibration}
\begin{figure}[t]
    \centering    {\includegraphics[width=.4\textwidth]{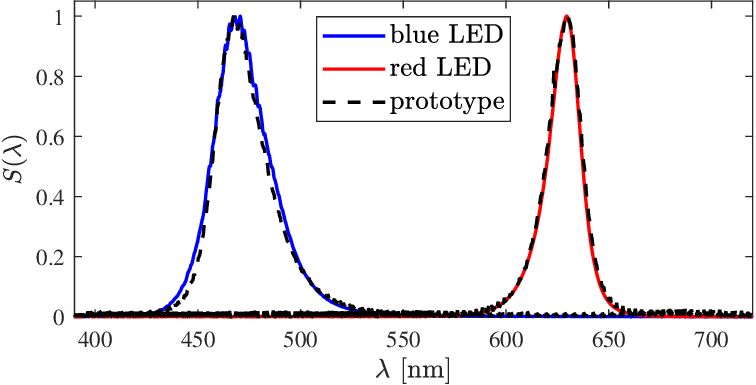}}
    \caption{Calibration of the $x$-axis of the prototype by comparing the measurements of two colored LEDs made by a commercial spectrophotometer..}
    \label{fig:calibration}
\end{figure}

The calibration of the position $s$ of the photosensors as a function of the wavelengths ($x$ scale of the SPDs) took place based on measurements from a commercial spectrophotometer model FLAME-T-XR1-ES from the manufacturer Ocean Optics, considering the peaks of the SPDs of two high-brigness LEDs with colors blue and red. In this way, and according the Fig. \ref{fig:calibration}, the wavelength of the first CCD sensor was adjusted to 391~nm and the last one to 723~nm. 
%\lcm{
Both calibration and measurements were performed in a temperature-controlled laboratory at 20$^\circ$C.
It is important to maintain this control, otherwise the PLA material may expand or contract, modifying the geometry of the structure at different temperatures.
%}
%\colr{(Discutir a accurácia do spectrophotometer construído. Como identificar/medir erros de calibração etc. )}

% -----------------------------------
\subsection{General description of operation}

Given what has been presented, the functioning of the meter can be summarized as follows:
\begin{enumerate} \setlength{\itemsep}{0pt}
    \item The light emitted by the luminaire to be analyzed is collimated by two pairs of parallel screens before reaching the diffraction grating;
    \item The diffraction grating decomposes the captured light, separating the wavelength bands in an angular way, generating a colored beam of light;
    \item The colored light beam reaches the photosensors of the CCD sensor which, in a scanning scheme, converts the intensities into analog voltage levels;
    \item The combination of digital synchronization and clock signals generated by \textmu C, switches the analog intensity signals of each sensor;
    \item \textmu C captures these signals using an ADC;
    \item These sampled signals are then sent to the PC, on which the CCT and CRI estimation algorithms are executed.
\end{enumerate}

% -----------------------------------
\subsection{Lighting quality analysis}

On the PC, the CCT and CRI are calculated according to the following steps:
\begin{enumerate}
\setlength{\itemsep}{0pt}
    \item In the intensity vector captured from the linear CCD sensor, dark calibration is performed, which consists of subtracting the average levels of the dark photosensors from the signals from the effective photosensors;
    \item To minimize distortions in the captured intensities as a function of $\lambda$, the inverse of \eqref{eq:curva_ccd} is applied, defining the SPD of the light source to be evaluated;
    \item The CCT is determined with \eqref{eq:ccta} or \eqref{eq:cctb};
    \item According to the CCT, a reference SPD is generated using blackbody radiation for values less than 5000~K (with \eqref{eq:refcorponegro}) or the standard illuminant D for values equal to or greater than 5000~K (with \eqref{eq:spdaprox}, \eqref{eq:m1} and \eqref{eq:m2}, \eqref{eq:xd1} or \eqref{eq:xd2}, and \eqref{eq:yd});
    \item The tristimulus values are calculated, and normalized to $Y=100$, for the evaluated illuminant, for the reference illuminant, and for the eight TCSs with \eqref{eq:X}, \eqref{eq:Y} and \eqref{eq:Z};
    \item Tristimulus values are transformed to the CIE 1931 color space (with \eqref{eq:x} and \eqref{eq:y}) and then to the CIE 1960 UCS uniform color space (with \eqref{eq:u1} and \eqref{eq:v1});
    \item Chromatic adaptation is applied to each sample with \eqref{eq:CA-c}, \eqref{eq:CA-d}, \eqref{eq:CA-u'ki} and \eqref{eq:CA-v'ki};
    \item The colorimetric information of the samples is transformed to the CIE 1964 color space with \eqref{eq:Wstar}, \eqref{eq:Ustar} and \eqref{eq:Vstar};
    \item The value $R_i$ is calculated for each sample with \eqref{eq:Ri}, through the colorimetric distances $\Delta E_i$ with \eqref{eq:deltaE};
    \item The mean of the $R_i$ values with \eqref{eq:CRI} is the CRI $\mathrm{R}_a$.
\end{enumerate}

% *********************************************
\section{Results} \label{sec:results}

Table \ref{tab:results} presents information and results for the CCT and CRI of fourteen light sources analyzed by the developed meter.
For comparison purposes, the nominal CCTs (defined by the manufacturer) of the lamps were identified.
In some cases, the manufacturer does not present this information on the product.
%\footnote{The fields in the Table \ref{tab:results} where the information was not found were filled with a hyphen.}.}
For all light sources analyzed, nominal CRI information was also not found. 
The fact that these important parameters are often neglected corroborates the need of the meter developed in this work. 

The biggest discrepancies were presented for the CCT, with a minimum, average, and maximum difference of 10.8\%, 24.6\% and 37.0\%. 
This can be attributed to considering the nominal value as a comparison, which may not be the most accurate, due to several factors, such as inconsistencies in the information provided by the manufacturer to the aging of the lamp.

For comparison purposes, the average values of twenty-nine fluorescent lamps, four incandescent lamps, two halogen lamps, and four LED lamps were considered in \cite{sandor2006,2004_cri} and \cite{2016_cri_congress}.
Regarding the CRI, the measured values were more coherent compared to the theoretical values, presenting a minimum, average, and maximum difference of 0.06\%, 3.2\% and 8.0\%, validating the developed meter.
As theoretically expected, halogen lamps, followed by incandescent lamps, have the highest CRIs.
LED-based lamps had slightly lower CRIs compared to the fluorescent lamps analyzed \cite{shaikh2018}.
It is also found that the LED flashlights of smartphones evaluated have a cold color temperature and lower color rendering indices when compared to the other LED lamps analyzed.

\begin{table*}[]
\centering
\caption{Prototype measurement results} 
\begin{tabular}{lcc|c|c|cc|cc}
\hline  
\multirow{2}{*}{Lamp type} & \multirow{2}{*}{Brand} & Power [W] & CIE 1931 & CIE 1960 &  \multicolumn{2}{c|}{${R}_a$ [\%]}  & \multicolumn{2}{c}{CCT [K]} \\ 
                 &         &  & $(x,\,y)$ & $(u,\,v)$ & measure   & \cite{sandor2006,2004_cri,2016_cri_congress}  & measure  & nominal  \\ \hline
\textbf{E27 halogen}	     & Ourolux  & 70  & (0.393,\,0.375) & (0.234,\,0.335) & 97.76    &  97,0    & 3703     & 2700     \\
\textbf{E27 incandescent} & Empalux  & 15  & (0.384,\,0.357) & (0.236,\,0.329) & 92.41   &  93.8    & 3775     & -   \\
\textbf{T8 fluorescent}   & G-light  & 36  & (0.342,\,0.347) & (0.211,\,0.321) & 90.67	&  82.7     & 4815     & 6400     \\
\hline
\textbf{E27 LED}          & Ourolux  & 6   & (0.284,\,0.285) & (0.194,\,0.292) & 89.90	&  \multirow{9}{*}{85.5}     & 7286 & 6500     \\
\textbf{T8 LED}           & Osram    & 18  & (0.312,\,0.354) & (0.189,\,0.321) & 88.98   &   & 5108	 & 6500     \\
{SPOT LED}  & Hoya     & 5  & (0.342,\,0.368)   & (0.203,\,0.328) & 82,84  &   & 5173	 & 4000    \\ 
{LED strip white} & -        & -  & (0.358,\,0.319)   & (0.203,\,0.328) & 79.32  &   & 4326	 & -    \\
{LED strip white warm} & -        & -  & (0.351,\,0.347)   & (0.217,0.322) & 88,78  &   & 4762	 & 3000   \\
{E27 LED}          & Onlux  & 4.9   & (0.294,\,0.316) & (0.189,\,0.306) & 83.55	&       & 7836 & 6500     \\
{E27 LED}          & Taschibra  & 9   & (0.359,\,0.356) & (0.219,\,0.326) & 89.74	&       & 4527 & 3000     \\
{E27 LED}          & Taschibra  & 4.9   & (0.297,\,0.335) & (0.185,\,0.313) & 83.00	&       & 7319 & 6500     \\
{E27 LED}          & Ourolux & 9   & (0.287,\,0.301) & (0.190,\,0.299) & 86.23	&       & 8790 & 6500     \\
\hline
{smartphone LED flashlight} & Xiami Redmi 12C & -   & (0.307,\,0.347) & (0.187,\,0.318) & 69.80	&  -     & 6673 & -     \\
{smartphone LED flashlight} & Motorola E7 plus & -   & (0.304,\,0.348) & (0.185,\,0.318) & 72.20	&   -   & 6816 & -     \\
\end{tabular}
\label{tab:results}
\end{table*}

\begin{figure}[t]
    \centering
    \includegraphics[width=.41\textwidth]{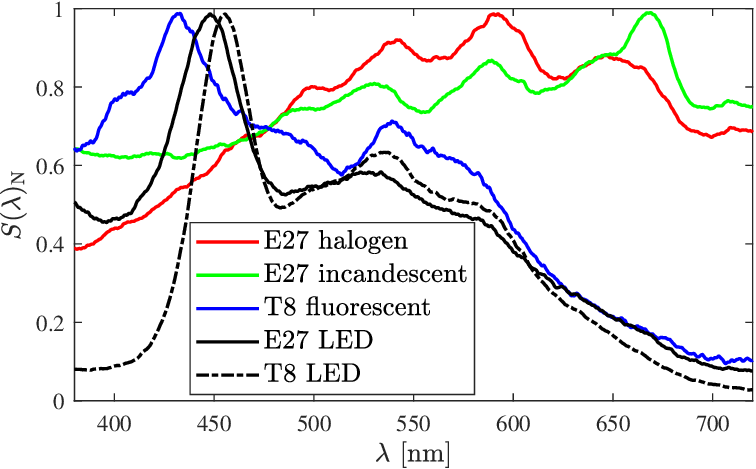}
    %\subfloat[]{\includegraphics[width=.4\textwidth]{fig/spda.eps}} \\
    %\subfloat[]{\includegraphics[width=.4\textwidth]{fig/spdb.eps}}
    \caption{SPDs measured by the prototype of the first five lamps in Table \ref{tab:results}.}
   \label{fig:spda_spdb}
\end{figure}

\begin{figure}[t]
    \centering
    \subfloat[]{\includegraphics[width=.323\textwidth]{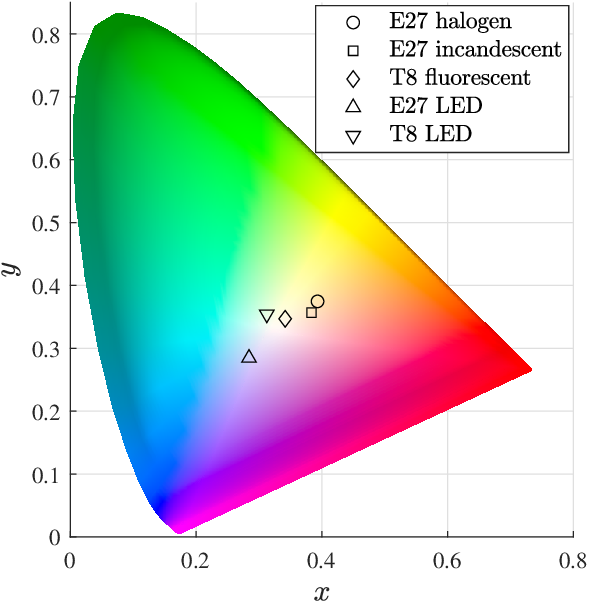}} \\ \vspace{-.2cm}
    \subfloat[]{\includegraphics[width=.43\textwidth]{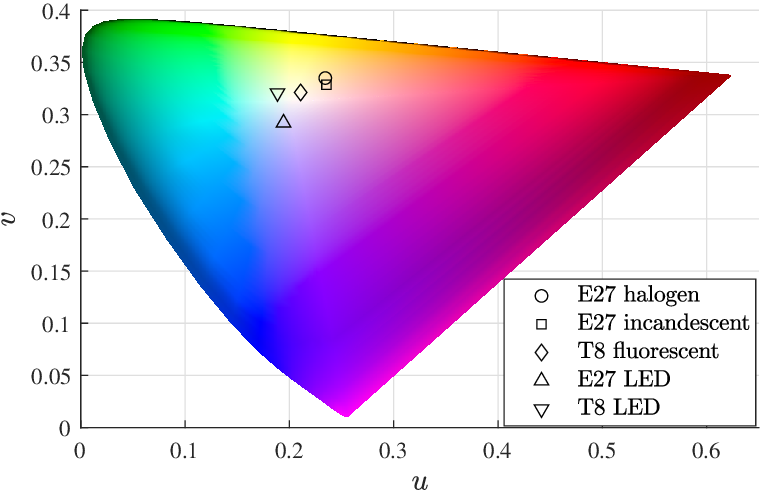}}
    \caption{Results for CIE 1931 color space in (a) and for CIE 1960 UCS color space in (b), for the first five lamps in Table \ref{tab:results}.}
    \label{fig:results}
\end{figure}

Fig. \ref{fig:spda_spdb} shows the measured SPDs, normalized by the peak. In order to not make visualization difficult, the results are presented for the first five light sources in Table \ref{tab:results}.
A similar shape to the blackbody spectrum is observed for the halogen lamp and the incandescent lamp.
It is also possible to notice the difference in format between the two LED lamps, with the warmer one (T8 LED) having relatively greater magnitudes in the red region compared to the E27 LED Ourolux 6~W.
In Figs. \ref{fig:results}(a) and \ref{fig:results}(b), the chromaticity coordinates of these lamps are printed to visually demonstrate the performance of each lamp.
The coordinates of the points in these two graphs confirm that the halogen and incandescent lamps have warmer colors (showing smaller CCTs), the T8 and fluorescent LEDs are more neutral, and the E27 LED Ourolux 6~W has the coldest color (higher CCT) among the luminaires considered.

% *********************************************
\section{Conclusion and Future Directions}\label{sec:concl}

This research successfully designed, constructed, and validated a low-cost, high-accuracy light fixture color temperature (CCT) and color rendering index (CRI) measuring device. The device utilizes a novel optical geometry, eliminating the need for expensive optical lenses and mitigating chromatic aberration, issues that are common in traditional spectrophotometers. This design significantly reduces cost and complexity while maintaining accuracy comparable to established commercial devices.

The prototype effectively captures spectral power distributions (SPDs) across a broad wavelength range. Linearity analysis demonstrated that the chosen optical arrangement provides suitable accuracy. The device underwent rigorous calibration using known SPDs from commercial LEDs verified by a commercial spectrophotometer. Subsequent measurements from various light sources, including halogen, incandescent, fluorescent, and LED, were used to evaluate the performance of the developed meter.

The results obtained for CRI values showed excellent agreement (within a maximum difference of 8.0\%) with established theoretical values and values reported in relevant literature. CCT measurements, while exhibiting higher deviations (reaching an average value of 24.6\%), are likely influenced by the use of nominal manufacturer values for comparison; more precise measurements could be made using calibrated standard lamps. 
Small discrepancies in the CCT may be attributed to the inaccuracy of the nominal values and variations between manufactured batches, among others.
%The slight discrepancies in CCT may be attributable to variations in lamp characteristics or minor calibration uncertainties. 
However, the consistency of the CRI values strongly validates the accuracy of the proposed spectrophotometric design and its implementation.

Looking ahead, there is a promising path for enhancing the system's calibration techniques to further boost CCT accuracy. 
Moreover, the software and display screen for displaying the results could be integrated with the developed spectrophotometer into a single stand-alone device, allowing greater versatility without needing an attached personal computer.
Additional testing and analysis of a wider range of applications, such as artificial lighting for agriculture, for example, could also strengthen the conclusions presented, instilling a sense of optimism about the future potential of the device.

This cost-effective and accurate device is not just a research tool, but a practical asset for the field of lighting quality assessment. It offers a reliable tool for researchers, manufacturers, and lighting designers, reassuring them of its usability. The methodology and results presented here lay a solid foundation for further advancements in low-cost, high-accuracy lighting metrology.

% The work presents the complete project of a lighting quality meter.
% In this sense, the appropriate arrangement of the components was modeled and determined based on two distinct geometric arrangements.
% An analysis of the opening of the collimators as a function of the expected resolution was included, as well as an analysis of the linearity of the photosensor positions as a function of wavelength.
% The latter verified the geometric arrangement with the sensor perpendicular to the average angle as more linear.

%The prototype required mechanical construction, electronic circuits, \textmu C, and software programming on the PC.
%After calibration, the SPDs of five different luminaires were captured to calculate the CRI and CCT.
%The CRI values obtained were consistent with those obtained in other references, validating the methodology and the proposed meter.

%\colr{(Quão consistente com outras referencias? Possível avaliar/calcular/discutir acurácia?)}

\color{black}
% use section* for acknowledgment
\section*{Acknowledgment}

%The authors would like to thank...
This work was supported in part by the National Council for Scientific and Technological Development (CNPq) of Brazil under Grant 314618/2023-6 and 405301/2021-9, and by the Coordination for the Improvement of Higher Education Personnel - Brasil (CAPES) - Finance Code 001. We also thank the physics laboratory at the Toledo campus of the Federal Technological University of Paraná (UTFPR).

% Can use something like this to put references on a page
% by themselves when using endfloat and the captionsoff option.
\ifCLASSOPTIONcaptionsoff
  \newpage
\fi

%\bibliographystyle{IEEEtran}
%%\balance
%\bibliography{refs.bib}

% Generated by IEEEtran.bst, version: 1.14 (2015/08/26)

\end{document}